\documentclass[epj,referee]{svjour}
\usepackage{graphicx, amssymb}
\begin{document}
\title{Fracture of a biopolymer gel as a viscoplastic disentanglement
process}
\author{Tristan Baumberger\mail{tristan@insp.jussieu.fr}, Christiane Caroli \& David Martina}
\institute{INSP, Universit\'e Pierre et Marie Curie-Paris 6,
Universit\'e Denis Diderot-Paris 7, CNRS, UMR 7588
Campus Boucicaut, 140 rue de Lourmel, 75015 Paris, France.}

\date{\today}

\abstract
{We present an extensive experimental study of
mode-I, steady, slow crack dynamics in gelatin gels. Taking advantage of the
sensitivity of the elastic stiffness to gel composition and history we
confirm and extend the model for fracture of physical hydrogels
which we proposed in a previous paper  (Nature 
Materials, doi:10.1038/nmat1666 (2006)),
which attributes decohesion to the viscoplastic pull-out of the
network-constituting chains. So, we propose that, in contrast with
chemically cross-linked
ones, reversible gels fracture without chain scission.}
\authorrunning{T. Baumberger {\it et al.}}
\maketitle

\section {Introduction}

Hydrogels are a family of materials constituted of a sparse random
polymer network swollen by a (most often aqueous) solvent. They can
be classified into two subgroups.

-- {\it Chemical gels}, such as polyacrylamid ones, in which the
cross-links (hereafter abbreviated as CL) between the polymer chains
are made of single covalent molecular brid\-ges. Their gelation process
is irreversible.

-- {\it Physical gels} in which cross-linking is due to hydrogen or
ionic bonds, much weaker than covalent ones. In most of them the
network is constituted of biopolymers \cite{Clark}, e.g. proteins (gelatin) or
polysaccharides (agar, alginates). Due to stabilizing steric
interactions, these CL may involve many monomeric units (residues),
extending over lengths of several nanometers. Such is the case for
gelatin gels. Gelatin results from the denaturation of collagen, whose
native triple helix structure is locally reconstituted in the CL
segments, interconnected in the gel by flexible segments of single
protein chains. Due to the weak strength of their CL bonds, physical
gels are thermoreversible. For example, gelatin networks "melt" close
above room temperature. This behavior leads to the well studied slow
ageing (strengthening) of their elastic modulus \cite{Nij}, and to their
noticeable creep under moderate stresses~\cite{HRM}.\\

Biopolymer based physical gels have been attracting increasing
interest motivated by their wide use in the food industry \cite{Vliet} and to
promising biomedical developments in fields such as drug delivery and
tissue engineering \cite{Lee}. All these implementations call for the control of
their mechanical properties -- namely elastic stiffness and fracture
toughness, independent tuning of which would be highly desirable.

While elastic responses of gels have been extensively studied, both
in the small \cite{Clark} \cite {Nij} and large deformation regimes 
\cite{Bot} \cite{Evoy}, fracture studies have been
up to now essentially concerned with crack nucleation \cite{Bonn} and ultimate
strength measurements \cite{Bot} \cite{Evoy}. However, trying to elucidate the nature of the
dissipative processes at play in fracture, which are responsible for
the rate dependence of their strength, naturally leads to
investigating the propagation of cracks independently from their
nucleation. Tanaka {\it et al} \cite{Tanaka} have performed such a study on 
chemical
polyacrylamid/water gels. By changing the concentration of cross-linking
agent at fixed polymer content, they found that, in
this material, stiffness and toughness are negatively correlated : as
is the case for rubbers, the stiffer the gel is, the smaller its
fracture energy. More recently, Mooney {\it et al} \cite{Mooney} 
have been able to compare
the fracture behavior of chemically and physically cross-linked
alginate gels. They showed that the stiffness/toughness correlation,
while agreeing with Tanaka's result for covalent CL, is inverted for
ionic ones. In this latter case "the stiffer the tougher".\\

We report here the results of an extensive study of steady, strongly
subsonic, mode-I
(opening) crack propagation in gelatin gels. This choice
was made for several reasons. First, due to their massive industrial
use, their elastic properties and molecular structures have been
thoroughly studied.
On the other hand, they can be easily cast into the large
homogeneous samples required for fracture experiments. Morevover,
solvent viscosity can be tuned by using glycerol/water mixtures.

We have studied the dependence of the fracture energy $\mathcal{G}$ on
the crack velocity $V$ for gels differing by their gelatin
concentration $c$, glycerol content $\phi$, and thermal history, each
of which is known to affect their elastic properties. Experimental
methods are described in Section 2. We present in Section 3. the
behavior of $\mathcal{G}(V)$ for $3$ different series of samples~:

A --- Common $c$ and history, variable $\phi$ (hence solvent viscosity
   $\eta_{s}$).
   
B --- Fixed $c$ and $\phi$, different histories.

C --- Common $\phi$ and history, variable $c$.

We discuss and interpret these results in Section 4. As already
reported in~\cite{Nature}, the analysis of solvent
effects (series A) leads us to propose that,
in contradistinction with chemical hydrogels, physical ones do not
fracture by chain scission, but by viscous pull-out of whole gelatin
chains from the network via plastic yielding of the CL. This
interpretation properly accounts for the quasi-linear dependence
of $\mathcal{G}$ on $\eta_{s}V$ as well as for the orders of
magnitude of its slope $\Gamma = d\mathcal{G}/d(\eta_{s}V)$ and of
its quasi-static limit $\mathcal{G}_{0}$. We then turn toward the
variations of $\Gamma$ with the small strain shear modulus $\mu^{*}$.
We find that our fracture scenario, when combined with the model
proposed by Joly-Duhamel {\it et al} \cite{JHAD}
for gelatin network structure and elasticity, is compatible with the
results from series B. One step further, the analysis of the effect
of gelatin concentration variations (series C) leads us to invoke a
concentation-dependent effective viscosity affecting the viscous drag
on chains pulled out of the gel matrix.

\section{Experimental methods}

\subsection{Sample preparation}
The gels are prepared by dissolving gelatin powder (type A from
porcine skin, 300 Bloom, Sigma) in mixtures containing a weight fraction $\phi$
of glycerol in deionized water, under continuous stirring for $30$ min
at 90$^\circ$C. This temperature, higher than commonly used ones 
($\sim$ 50 - 60$^\circ$C) has been chosen, following Ferry 
\cite{Ferry}, so as to obtain homogeneous pre-gel solutions even at the highest
$\phi$ (60 $\%$). A control experiment carried out with a (pure
water)/gelatin sample prepared at 60$^\circ$ C resulted in
differences of low strain moduli and $\Gamma$ values of, respectively,
1 $\%$ and 7 $\%$, compatible with scatters between 90$^\circ$C
samples. So, we concluded that our preparation method does not, as
might have been feared, induce significant gelatin hydrolysis.

The pre-gel solution is poured into a mould consisting of a
rectangular metal frame and two plates covered with Mylar films. On
the longest sides of the frame, the curly part of an adhesive Velcro
tape improves the gel plate grip. Unless otherwise specified (see
Section 3.2, series B results), the thermal history is fixed as
follows. The mould is set at 2$\pm$0.5$^\circ$C  for 15 h, then clamped to the
mechanical testing set-up and left at room temperature (19$\pm$1 $^\circ$C)
for 1 h. This waiting time ensures that variations of elastic
moduli over the duration of the subsequent run can be safely 
neglected \cite{Nij}.
The removable pieces of the mould are then taken off, leaving the
300$\times$30$\times$10 mm$^{3}$ gel plate fixed to its grips. The
Mylar films are left in position to prevent solvent evaporation. They
are peeled off just before the experiment.

\subsection{Gel characterization}
\begin{figure}[h]
    \centering
    \includegraphics[scale=0.3]{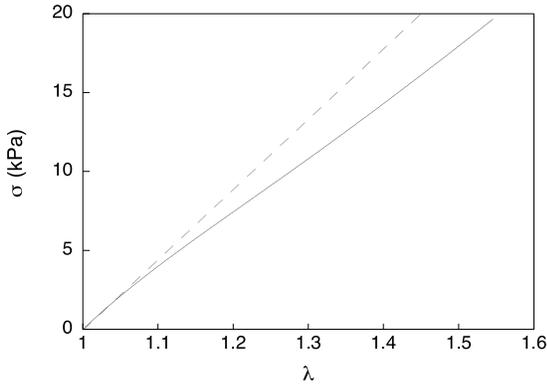}
    \caption{Nominal stress $\sigma = F/(e_{0}L_{0}$) versus stretching 
    ratio for a $c = 10$ wt\%, $\phi = 0$ wt\% sample plate. The dashed line is the 
    extrapolation of the small strain linear response. Its slope is 
    four times the effective shear modulus $\mu^{*}=11$ kPa (see text).}
    \label{fig:load}
\end{figure}

For each fracture experiment we prepare simultaneously two nominally
identical samples, one of which is used to determine the elastic
characteristics. For this purpose, with the help of the mechanical set
up described below, we measure the the force-elongation response
$F(\lambda)$ of
the plate (see Fig.~\ref{fig:load}), up to stretching ratios $\lambda$ = 1.5, at the loading
rate $\dot{\lambda}$ = 1.7 10$^{-2}$ sec$^{-1}$.

From these data, we extract an effective small strain shear modulus
$\mu^{*}$. In hydrogels, while shear stresses are sustained by the
network, pressure is essentially borne by the solvent. Hence, since
shear moduli are typically in the 1 - 10 kPa range, the gels can be
considered incompressible (Poisson ratio $\nu$ = 1/2), as long as no
solvent draining occurs \cite{DLJ}. So,  the  sound velocity
relevant to define the subsonic regime is the transverse one $c_{s} =
\sqrt{\mu/\rho}$, with $\rho$ the gel mass density. For our systems,
typically $c_{s} \sim 1$ m.sec$^{-1}$.
Neglecting finite size effects, we assume plane
stress uniform deformation for our plates of undeformed length
$L_{0}$ = 300 mm, width $h_{0}$ = 30 mm, thickness $e_{0}$ = 10 mm. In
the linear regime, this assumption leads us to define a (necessarily
somewhat overestimated) effective modulus as $\mu^{*} = \frac{1}{4}
\left(\frac {d\sigma}{d\lambda}\right)_{\lambda = 0}$, with $\sigma =
F/(e_{0}L_{0})$ the nominal stress, $\lambda = h/h_{0}$ the stretching
ratio, $h$ the stretched width.

One step further, and under the conservative assumption that small
strain elasticity is basically of entropic origin, we extract a length
scale characteristic of the network as $ \xi =
\left(k_{B}T/\mu^{*}\right)^{1/3}$, which lies in the 10 nm range.
This order of magnitude agrees with the one which can be evaluated
from measurements of the collective diffusion coefficient $D_{coll}$
which characterizes the solvent/network relative motion 
\cite{DLJ} \cite{THB}.

For gelatin/water samples \cite{Ronsin}, $D_{coll} \sim 10^{-11}$
m$^{2}$/sec, so that a typical time scale for draining over $\sim 1$cm
is on the order of $10^{7}$ sec, which means that macroscopic
stress-induced draining is totally negligible here.\\

As can be seen on Figure~\ref{fig:load}, beyond $\lambda$ values on the order of
1.1, the force response markedly departs from its small strain linear
behavior. In order to calculate the mechanical energy released per
unit area of crack extension, conventionally termed energy release
"rate" $\mathcal{G}$, we need to compute the elastic energy
$\mathcal{F}(\lambda)$ stored in the stretched plate. For this
purpose we integrate numerically the measured response curve.

\subsection{Fracture experiments}
\begin{figure}[tbp]
    \centering
    \includegraphics[scale=0.5]{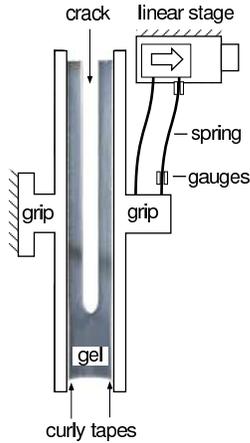}
    \caption{Schematic representation of the mechanical setup, drawn 
    around a genuine photograph of a gel plate ($c = 10$ wt\%, 
    $\phi = 0$ wt\%), stretched to 
    $\lambda = 1.5$. Note that the crack propagates 
    straight along the mid-plane. The light blue hue of the gel (color 
    on line) results from Rayleigh scattering by small scale gel 
    network randomness.}
    \label{fig:setup}
\end{figure}

The mechanical set-up is sketched on Figure~\ref{fig:setup}. One of the grips
holding the gel plate is clamped to the rigid external frame. The
other one is attached to one end of a double cantilever spring of
stiffness $K = 43.1\times10^{3}$ N.m$^{-1}$. The other end of the spring
can be displaced by a linear translation stage, with a $0.1 \mu$m
resolution. The deflection of the spring is measured by four strain
gauges glued to the spring leaves, with a resolution of $5.10^{-2}
\mu$m.

In most runs, the sample stiffness is much smaller than the spring
one, and fracture occurs in the so-called fixed grips configuration.
The stretching ratio $\lambda$ is computed in all cases by
subtracting the spring deflection amplitude from the stage
displacement.

Before stretching, a knife cut of length 20 mm is made at mid-width at
the upper free gel edge. In a first set of
experiments the grips are pulled apart for 1 sec up to the desired
amount $\Delta h$. The resulting crack advance is monitored by a
camera with a 631 $\times$ 491 pix$^{2}$ CCD device operating at a
typical rate of 15 sec$^{-1}$. The crack tip position is measured with
0.5 mm resolution. The crack velocity $V$ is obtained from a sliding
linear regression over 5 successive position data.

Away from the sample edges, in this configuration, cracks run at
constant velocity \footnote{ This is true for not too small
velocities, where bulk creep during a run is negligible. For slow
cracks, with velocities below a few hundred $\mu$m.sec$^{-1}$, creep
results in a measurable velocity drift. We only retain data out of
this range.}. As expected, the free edges affect crack propagation up
to a distance comparable with the plate width. Further data processing
has been systematically restricted to the central region, extending
over $\sim$ 200 mm. In this region, we can legitimately compute the
energy release rate as \cite{Rivlin}
$\mathcal{G} = \mathcal{F}/(e_{0}L_{0})$.

Such experiments result in one run producing one single $\mathcal{G} -
V$ data point, hence are very time consuming. So, in a second set of
experiments, the stretching ratio was increased at the constant rate
$\dot{\lambda}$ = 1.7 10$^{-2}$ sec$^{-1}$. This results in a slowly
accelerating crack. We have validated the corresponding
$\mathcal{G}(V)$ data by comparison with steady state ones on an
overlapping velocity range (see Fig.~\ref{fig:G(V)}). 
The crack dynamics in this latter type of
experiments can therefore be termed "quasi-stationary".

\section{Experimental results}

\subsection{Solvent effects}
We summarize here the results, already reported in
reference \cite{Nature}, corresponding to series A, namely gels
prepared as described above, with gelatin concentration $c$ = 5
wt$\%$, glycerol content ranging from $0$ to 60 wt$\%$, i.e. solvent
viscosity $\eta_{s}$ from 1 to 11 times that of pure water.

\begin{figure}[h]
    \centering
    \includegraphics[scale=0.3]{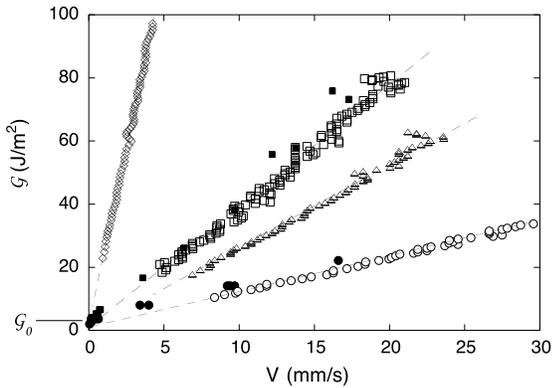}
    \caption{Fracture energy release rate for gels with the same 
    gelatin concentration ($c = 5$ wt\%) and various glycerol contents 
    (series A):
    $\phi$ = 0 wt\% (circles), 20 wt\% (triangles), 30 
    wt\% 
    (squares), 60 wt\% (diamonds). Filled symbols correspond to 
    stationary cracks,  open symbols to 
    cracks accelerated in response to a steady increase of $\lambda$. 
    $\mathcal G_{0} = 2.5 \pm 0.5$  
    J.m$^{-2}$ is the common linearly extrapolated toughness. From 
    ref.~\cite{Nature}. 
    (reprinted from Nature Materials).}
    \label{fig:G(V)}
\end{figure}

As shown on Figure~\ref{fig:G(V)}, for all samples $\mathcal{G}$ increases
quasi-linearly with $V$ in the explored range and, within experimental
accuracy, the various curves extrapolate to a common,
$\phi$-independent value $\mathcal{G}(V \rightarrow 0) =
\mathcal{G}_{0}$ which yields an evaluated quasi-static toughness.
This cannot be accessed directly. Indeed, the above mentioned
importance of creep in our gels leads to the well-known problems met
when trying to define static threshold in weak solids (such as
colloidal gels, pastes,\ldots). For this series, we find
$\mathcal{G}_{0} \simeq$ 2.5 J m$^{-2}$, a value about 20 times smaller
than a gel-air surface energy.

\begin{figure}[h]
    \centering
    \includegraphics[scale=0.3]{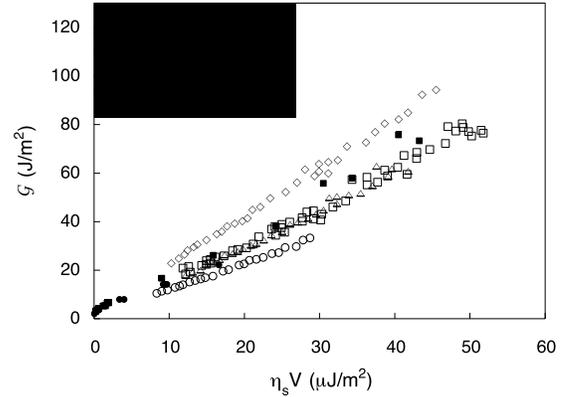}
    \caption{Same data as Fig.~\ref{fig:G(V)} replotted versus 
    $\eta_{s}V$, with $\eta_{s}$ the viscosity of the glycerol/water 
    solvent. From 
    ref.~\cite{Nature}. 
    (reprinted from Nature Materials).}
    \label{fig:G(etaV)}
\end{figure}

\begin{figure}[h]
    \centering
    \includegraphics[scale=0.3]{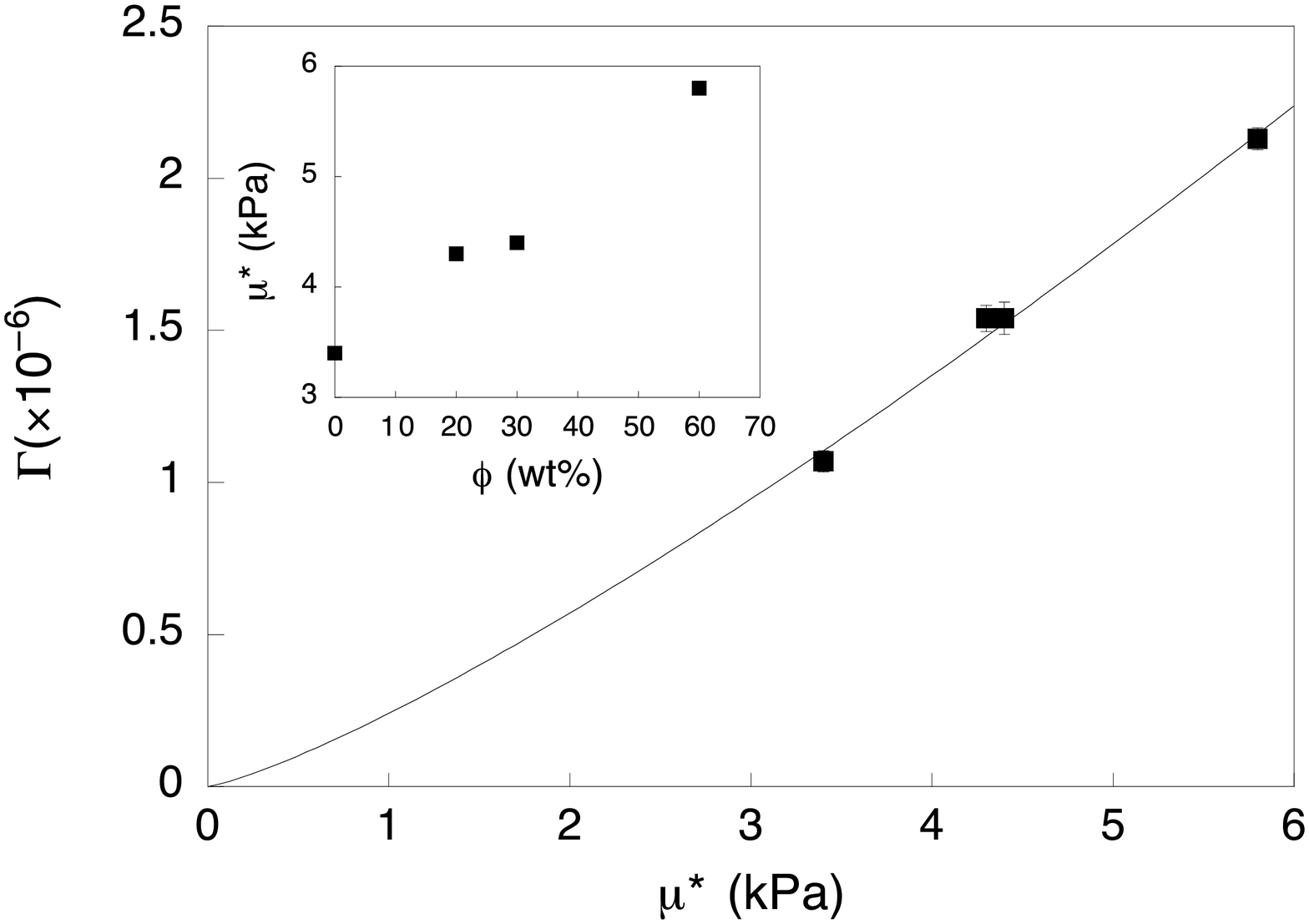}
    \caption{Rate sensitivity $\Gamma = d{\mathcal G}/d(\eta_{s}V)$ 
    vs. $\mu^{*}$ for the samples of series A. The line is the best power law 
    fit $\Gamma\sim \mu^{*1.2}$. Insert shows that increasing the 
    glycerol content stiffens the gel.}
    \label{fig:Gamma(mu)A}
\end{figure}

Moreover, the slope $d\mathcal{G}/dV$ strongly increases with $\phi$,
which suggests that $\eta_{s}V$ might be the relevant variable.
Indeed, the corresponding plot (Fig.~\ref{fig:G(etaV)}) captures most of this
variation. We therefore write
\begin{equation}
\label{eq:gammadef}
\mathcal{G} = \mathcal{G}_{0} + \Gamma \,\eta_{s}V
\end{equation}
The dimensionless slope $\Gamma$ is found to be a huge number, of
order $10^{6}$. In Section 4 below, we
will relate the variations of $\Gamma$ with those of the elastic
modulus $\mu^{*}$. Figure~\ref{fig:Gamma(mu)A} shows that, within series A, $\Gamma$
increases with $\mu^{*}$.

\begin{figure}[h]
    \centering
    \includegraphics[scale=0.3]{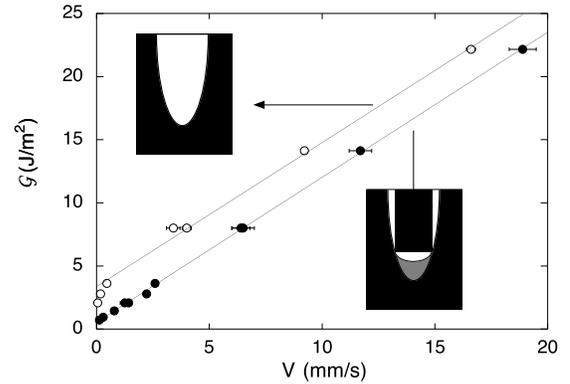}
    \caption{$\mathcal G(V)$ curves for a 5 wt\% gelatin gel in pure 
    water~:  ``dry" 
    cracks opening in 
    ambient air (upper data) and ``wet" 
    cracks with a drop of pure water soaking the tip. 
    At $\mathcal G$ too low  for dry cracks to 
    propagate, wet ones can still run. Linear fits 
    are shown. The wet data appear merely translated towards lower energies.
    The extrapolated fracture energy for wet tips is $\mathcal 
    G_{0}^{wet} = 0.6 \pm 0.15$ J.m$^{-2}$.
    From 
    ref.~\cite{Nature}. 
    (reprinted from Nature Materials).}
    \label{fig:Gdry/wet}
\end{figure}

The quasi-scaling of $\mathcal{G}$ with $\eta_{s}V$ points toward the
critical role of polymer-solvent relative motion in the fracture
process. In order to shed further light on this point, we have also
performed, with the same gels, experiments in which a small drop of
solvent is introduced into the already moving crack opening. For such
wetted cracks, as shown on Figure~\ref{fig:Gdry/wet}, $\mathcal{G}(V)$ is simply
shifted downward by a constant amount $-\Delta\mathcal{G}_{0}$, its
slope remaining unaffected. The energy cost $\Delta\mathcal{G}_{0} \sim
$ 2 J m$^{-2}$, a substantial fraction of $\mathcal{G}_{0}$. It
clearly signals that, in the non-wetted tip case, fracture involves
exposing gelatin chains to air. Such local solvent draining into the
gel bulk is likely to result from the impossibility for our not very
thin incompressible plates to accommodate the high strain gradients which
develop close to the tip without being the seat of high negative fluid
pressures.

In a static situation, the solvent would get sucked from the bulk into the tip
region, leading to  
gradual smearing out of the fluid
pressure gradient. However, in the steadily moving case, the space range of this
collective diffusion process is limited to $\sim D_{coll}/V$ 
\cite{Rud} \cite{Ruina}.
For tip velocities above $\sim 1$ mm sec$^{-1}$, this length is smaller
than the mesh size $\xi$, and the process is inefficient. For much
slower cracks, it would lead to a long transient towards a lower
apparent $\mathcal{G}_{0}$. Trying to disentangle this from creep
effects, which also become relevant for slow cracks, will demand a
detailed characterization of creep which is out of the scope of this
paper.

\subsection{History-controlled stiffness effects}
\begin{figure}[h]
    \centering
    \includegraphics[scale=0.3]{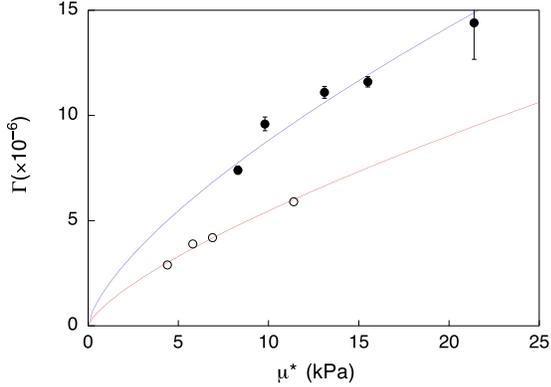}
    \caption{$\Gamma$ vs. $\mu^{*}$ for   gels from series B ($\phi = 
    0$, various thermal histories). 
    $c = 15$ wt\% (full dots); $c = 10$ wt\% (open circles).  The curves are 
    guide for the eye.}
    \label{fig:Gamma(mu)B}
\end{figure}

The results for series A above suggest a positive correlation between
the slope $\Gamma$ and the small strain modulus $\mu^{*}$. In a second
set of experiments, we have tuned $\mu^{*}$ at two different gel
compositions, namely $\phi$ = 0, $c$ = 10 and 15 wt$\% $. This was
realized by taking advantage of the rather strong dependence of
$\mu^{*}$ on the temperature maintained during gelation, as well as on
the duration of the gelation phase itself \cite{Nij} \cite{JHAD}
(always chosen large enough for $\mu^{*}$ variations to remain negligible during the run).
This enabled us to induce $\mu^{*}$ values differing by at most a
factor of 2. The data are shown on Figure~\ref{fig:Gamma(mu)B}. It is seen that,
for each $c$-value, again, the stiffer the gel, the tougher. Note, however,
that $\Gamma$ is not a function of $\mu^{*}$ only, but also of
composition - a point which will be discussed in detail in Section 4.

\subsection{Gelatin concentration effects}
\begin{figure}[h]
    \centering
    \includegraphics[scale=0.3]{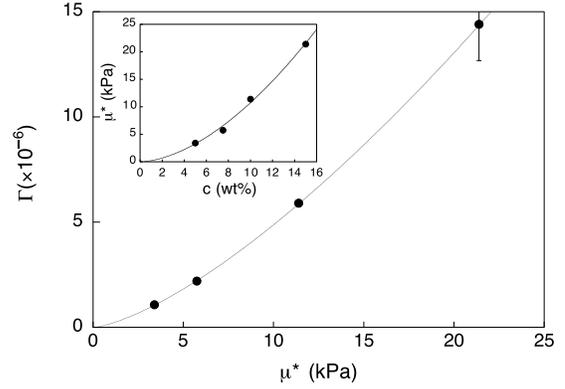}
    \caption{$\Gamma$ vs. $\mu^{*}$ for  gels from series C ($\phi = 
    0$, various gelatin concentrations). Insert shows $\mu^{*}$ vs. 
    $c$. The full lines are the power law fits (see text).}
    \label{fig:Gamma(mu)C}
\end{figure}

We have investigated this last point directly by working with a third
set of samples (series C) with the common history described in
section 2, the same solvent (pure water) and different values   of
$c$. As already amply documented \cite{Clark} \cite{JHAD}, 
$\mu^{*}$ increases with $c$ (Fig.\ref{fig:Gamma(mu)C}). 
A power law fit yields $\mu^{*} \sim c^{1.64 \pm 0.2}$. This
exponent, somewhat lower than usual values ($\lesssim 2$), is close to
that measured by Bot {\it et al} \cite{Bot}. 
Figure~\ref{fig:Gamma(mu)C} also shows the $\Gamma(\mu^{*})$
data. Once more, $d\Gamma/d\mu^{*} > 0$.

\section{Discussion and interpretation}

\subsection{A viscoplastic model of gelatin fracture}
At first glance, as far as fracture is concerned, our gels share two
salient features with another class of soft elastic materials, namely
rubbers \cite{LT} \cite{Knauss}. In both cases :

   (1) the toughness $\mathcal{G}_{0}$ is at least one order of
   magnitude larger than the energy of the surfaces created by the crack
  advance.

   (2) $\mathcal{G}$ increases rapidly with $V$ in the strongly subsonic regime.

Hence a first question : are the physical mechanisms now well
established to be responsible for these two features in the case of
rubbers also at work for our physical gels?

The basic theory of rubber toughness was formulated by Lake and
Thomas \cite{LT}. Fracture occurs via chain scission : the polymer segments,
of areal density $\Sigma$, crossing the fracture plane are
stretched taut until they store an elastic energy per monomer on
the order of the covalent monomer-monomer bond one, $U_{chain} \sim$
a few $eV$. At this stage, each of them sustains a force $f_{chain}
\sim U_{chain}/a$, with $a$ a monomer size. A bond-breaking event
thus corresponds to dissipating all of the elastic energy that was
stored in the whole segment ($n$ monomers) joining two cross-links,
$\sim nU_{chain}$. So, $\mathcal{G}_{0}^{(rub)} \sim
nU_{chain}\Sigma$, an expression which explains the order of
magnitude of $\mathcal{G}_{0}$ as well as its decrease when stiffness
increases (the stiffer a rubber is, the less tough).

The $V$-dependent fracture energy of rubbers is of the form \cite{GS} 
\cite{Gent}

\begin{equation}
\label{eq:rubber}
\mathcal{G}^{(rub)} (V) = \mathcal{G}_{0}^{(rub)} [1 + \Phi(a_{T}V)]
\end{equation}
where $a_{T}$ is a temperature dependent WLF-like factor. This
velocity dependence has been shown to result from bulk viscoelastic
dissipation \cite{PGG} \cite {HXK}. 
Due to the stress gradients ahead of the moving crack,
which extend far beyond the "active tip zone" where decohesion takes
place, the material deforms at a strain rate which
sweeps its whole relaxation spectrum, hence the WLF scaling factor.
That $\mathcal{G}_{0}^{(rub)}$ factors out in expression
(\ref{eq:rubber}) results from two facts \cite{BB} : 
(i) linear elasticity
preserves the universal $r^{-1/2}$ stress concentration field (ii)
the so-called small scale yielding assumption holds, namely the size
of the active zone is negligible as compared with that of the viscous
dissipating one.\\

We will now argue that none of these mechanisms is relevant in our case.

On the one hand, we claim that, in physical gels, fracture cannot
process via chain scission. Indeed, the force $f_{chain}$ defined
above is more than one order of magnitude larger than that, $f^{*}
\simeq U_{CL}/a$, which can be sustained by the H-bond stabilized
cross-links. Clearly, when the stored elastic energy reaches $\sim U_{CL}$
per monomer, CL bonds yield, by either unzipping \cite{Ball} 
\cite{Nishi} or frictional sliding \cite{Bo}.
This leads us to postulate that, in the highly stressed active tip
zone, the chains which cross the crack plane creep until they are fully
pulled out of the gel matrix. The threshold stress at the
onset of CL yielding is $\sigma^{*} = f^{*}\Sigma$, with $\Sigma$ the
areal density of crossing chains. As a rough estimate for this density
we take $\Sigma \sim 1/\xi^{2}$, with

\begin{equation}
\label{eq:xi}
\xi = \left(\frac{k_{B}T}{\mu^{*}}\right)^{1/3}
\end{equation}
the above-defined estimate of the mesh size of the polymer network.
Then, with $a \sim 0.3$nm, $U_{CL} \sim 0.1$eV, $\xi \sim 10$nm, we
obtain $\sigma^{*} \sim 500$kPa.

Note that, contrary to standard conditions met with hard materials,
here $\sigma^{*}/\mu^{*} \gg 1$ ($\sim 10^{2}$), which makes the
issue of elastic blunting raised by Hui {\it et al} \cite{Blunt} 
certainly relevant to
gel fracture.

When solvent can be pumped from a wetting drop (see Section 3.1),
the plastic zone deforms under this constant stress until the
opening $\delta_{c}$ at the tip reaches the length of the chain - i.e.
its full contour length $l$, since at this stress level it is pulled taut.
This is precisely the well-know Dugdale model of fracture \cite{Lawn}, which
yields, for the quasi-static fracture energy of wet cracks :
\begin{equation}
\label{eq:wet}
\mathcal{G}_{0}^{wet} = \sigma^{*} l
\end{equation}
From series A resuts, we estimate $\mathcal{G}_{0}^{wet} \approx 0.6
\pm 0.15$ J m$^{-2}$. This, together with expression (\ref{eq:wet}),
enables us to get an estimated chain contour length $l \sim 1.2 \mu$m.
With an average mass $M_{res} = 80$ g/mole for each of the $l/a$
residues, this means a reasonable $300$ kg order of magnitude estimate
for the gelatin molar weight.

In this picture, we interpret the shift $\Delta\mathcal{G}_{0} =
\mathcal{G}_{0} - \mathcal{G}_{0}^{wet}$ as an energy cost associated
with chain extraction out of the solvent. This yields for the
solvation energy per chain $\Delta\mathcal{G}_{0}\xi^{2} \sim 1000$
eV, i.e. $\sim 10 k_{B}T$ per residue.

Let us now turn to the $V$-dependence of $\mathcal{G}$. The tip
wetting experiments (see Figure~\ref{fig:Gdry/wet}) directly show that
$\mathcal{G}_{0}$ and the slope $\Gamma$ are independent : wetting
shifts $\mathcal{G}_{0}$ while leaving $\Gamma$ unaffected. We
consider that this empirical argument by itself rules out bulk
viscoelasticity as the controlling mechanism. This appears
all the more reasonable that rheological studies \cite{Nij} 
\cite{Ferry} show that viscous
dissipation in hydrogels (loss angles typically $\lesssim 0.1$) is
much smaller than that in rubbers.

We are therefore led to extend our fracture model to finite velocities.
A finite $V$ means  a finite average pull-out velocity
$\dot{\delta} = \alpha V$, where $\alpha$ is a geometrical factor
characteristic of the shape of the Dugdale zone. Pull-out implies
motion of the network relative to the solvent, hence a viscous
contribution to the viscoplastic tip stress :
\begin{equation}
\label{eq:sigma}
\sigma_{tip} = \sigma^{*} + \sigma_{vis}(V)
\end{equation}
Solvent/network relative motion is diffusive \cite{DLJ}, 
which implies that fluid
pressure gradients obey a Darcy law with an effective porosity
$\kappa = \eta_{s} D_{coll}/\mu$, which can be expected on dimensional
grounds to scale as $\xi^{2}$. Baumberger {\it et al} \cite{Ronsin} 
have shown that, for
gelatin gels such as used in this work, $\kappa/\xi^{2} \simeq
6.10^{-2}$. We thus estimate $\sigma_{vis}$ as resulting from the
build up of the Darcy pressure over a length $\sim l$, i.e.
\begin{equation}
\label{eq:sigvis}
\sigma_{vis} \sim  l\,(\nabla p)_{Darcy} \sim
\frac{l \eta_{s}\dot{\delta}}{\kappa}
\end{equation}
and
\begin{eqnarray}
 \label{eq:vis}
 \mathcal{G}(V)&\approx &\mathcal{G}_{0} + l \sigma_{vis}\nonumber\\
 &= &\mathcal{G}_{0} + \alpha \frac{l^{2}}{\kappa} \eta_{s}V
 \end{eqnarray}
 which exhibits the observed linear variation with $\eta_{s}V$ and
 predicts that the slope
 \begin{equation}
 \label{eq:gamma}
 \Gamma = \alpha\frac{l^{2}}{\kappa}
 \end{equation}

 We found (Section 3.1) that $\Gamma$ is of order $10^{6}$. With $l$
 as evaluated above and $\xi \sim 10$ nm, we get from expression
 (\ref{eq:gamma}) $\Gamma \approx 2. 10^{5} \alpha$, which suggests
 that $\alpha$ should be of order $1$ at least. In the Dugdale model,
 one gets :
 \begin{equation}
 \label{eq:alpha}
 \alpha  = \frac{\delta_{c}}{d_{act}} \approx \frac{\sigma^{*}}{\mu}
 \end{equation}

For hard solids, $\sigma^{*}$ is the plastic yield stress
$\sigma_{Y}$, always~$\ll \mu$. We pointed out that, for physical
 gels, on the contrary, $\sigma^{*}/\mu \gg 1$. The Dugdale analysis
 can certainly not be directly used here, due to the very large
 deformation levels involved, hence to problems such as elastic
 blunting, strain-hardening and strain induced helix-coil transitions
 \cite{Courty}. We were able, with the help of a hetero-wetting
 experiment (pure water wetting a crack tip in a glycerolled gel)
 reported in \cite{Nature}, to obtain a direct evaluation of the size
 of the active zone. It yielded $d_{act} \sim 100$ nm, from which we
expect that $\alpha = l/d_{act} \sim 10$.

We should point out that our
model for tip dissipation (Eq.~(\ref{eq:sigma})) is formally
identical to that put forward by Raphael and de Gennes \cite{Elie} 
in the context
of rubber-rubber adhesion with connector molecules. But in the gel case,
where viscous dissipation is controlled by solvent-network friction,
the very large compliances involved cast doubt on the legitimity of mathematical
treatments based upon small opening and linear elasticity
approximations \cite{Elie} \cite{Fager}. However, the possibility of accessing
$d_{act}$, and thus the fracture parameter $\alpha$ experimentally,
together with the absence of substantial bulk viscoelastic dissipation
enable us  to conclude that our fracture model is consistent with experiments
as far as :

  -- it accounts for the linear dependence of $\mathcal{G}$ on
  $\eta_{s}V$.

  -- it yields reasonable orders of magnitude for the quasi-static
  toughness and the slope $\Gamma$.

\subsection{Relationship between fracture and elastic properties}

For further confirmation we now need to test the predictions of the
model against the measured variations of $\Gamma$ with small strain
elastic modulus $\mu^{*}$.

\begin{figure}[h]
    \centering
    \includegraphics[scale=0.3]{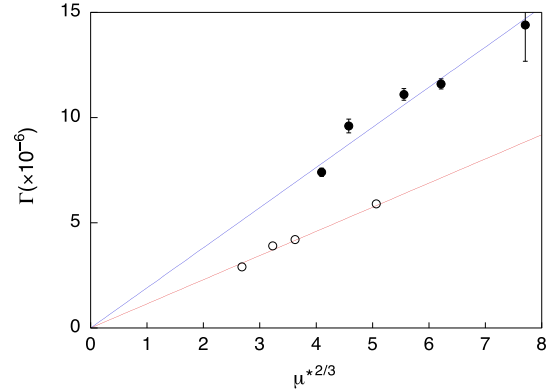}
    \caption{Data from Fig.~\ref{fig:Gamma(mu)B} replotted versus 
    $(\mu^{*})^{2/3}$ (eq.~(\ref{eq:scale})).}
    \label{fig:Gamma(mu2/3)B}
\end{figure}

Let us first consider the results of series B, involving gels with the
same composition but various thermal histories. According to
equation (\ref{eq:gamma}) we predict that, for each such set of
samples, $\Gamma$ should scale as  $\kappa^{-1}$, i.e. as :
\begin {equation}
\label{eq:scale}
\Gamma \approx \mu^{2/3}
\end{equation}
As seen on Figure~\ref{fig:Gamma(mu2/3)B}, the agreement with experimental data is quite
satisfactory, bringing good support to the model.

Note, however, that the two data sets pertaining to the two different
gelatin concentrations do not collapse onto a single master curve
(here a straight line). That is, the fracture "rate sensitivity"
$\Gamma$ does not depend on one single structural parameter. This
remark must be considered in the light of the finding by Joly-Duhamel
{\it et al} \cite{JHAD} (hereafter abbreviated as JHAD) that, for gels of various
gelatin concentrations, glycerol contents and thermal histories, there
is a one-to-one correspondence between the storage modulus $\mu$ and
the so-called helix concentration $c_{hel}$. This latter structural
parameter, directly obtained from optical activity measurements, is
interpreted as proportional to the length of triple-helix cross-links
per unit volume of gel. One might then be tempted to think that the
modulus $\mu$ contains essentially all the mecano-structural
information about the gel. That such is not the case is shown by two
observations :

  (i) JHAD also show that the loss modulus $\mu''$ does not depend on
  $c_{hel}$ only, but also on e.g. the gelatin concentation $c$.

  (ii) A non universal behavior was also found by Bot {\it et al} 
  \cite{Bot} for the
  non-linear part of the stress response in compression and in
  shear - a result confirmed by our own data.\\

\begin{figure}[h]
    \centering
    \includegraphics[scale=0.3]{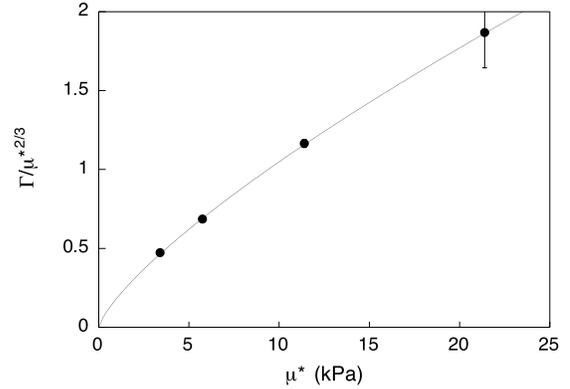}
    \caption{Data from Fig.~\ref{fig:Gamma(mu)C} replotted as 
    $\Gamma/(\mu^{*})^{2/3}$  vs. $\mu^{*}$. The line is the best power 
    law fit (exponent $0.75$).}
    \label{fig:Gamma/mu2/3C}
\end{figure}

We therefore now turn to the results of series C, which involve gels
with the same history and glycerol content ($\phi = 0$) and four
different values of $c$. As can be seen on Figure~\ref{fig:Gamma/mu2/3C},
$\Gamma/(\mu^{*})^{2/3}$ definitely increases with $\mu^{*}$, i.e. with
gelatin concentration. It was shown in JHAD that, in the range of
moduli explored here ($\mu > 2$ kPa), gel elasticity is well described as
that of a freely-hinged network of triple helix rods with average
distance $d \sim (k_{B}T/\mu)^{1/3}$, i.e. scaling as the mesh
length scale $\xi$. This leaves the $\kappa^{-1 } \sim \mu^{2/3}$ scaling
unaffected. We are thus led to attributing the residual variation of
$\Gamma$ to a concentration dependence of the viscosity appearing in
the poroelastic Darcy law. We propose that this should involve, not
the bare solvent viscosity, but an effective one 

\begin{equation}
\label{eq:etaeff}
\eta_{eff}(c) = \eta_{s}\,\Theta(c)
\end{equation}

including possible contributions from dangling ends, loops attached
to the network or free chains, invoked in JHAD and in Tanaka's study 
\cite{Tanaka}
of the fracture of chemical gels. 
In view of the discussion (see Section 4.1) of the order of
magnitude of $\Gamma$, clearly, $\Theta(c)$ should be $\mathcal{O}(1)$.

A tentative power law fit (Figure~\ref{fig:Gamma/mu2/3C}) yields 
$\eta_{eff}(c) \sim
(\mu^{*})^{0.75\pm0.03}$ which, combined with the $\mu^{*}(c)$ variations 
(see section 3.3), results in $\eta_{eff}(c)/\eta_{s}\sim c^{1.2}$. 
The study of creep viscosity in gelatin by Higgs and
Ross-Murphy \cite{HRM} concluded to a $c^{1.1}$ variation. However, their work
was concerned with stress levels  ($\sigma/\mu$ from $2.10^{-2}$ to
$2.10^{-1}$) considerably smaller than those relevant to the active
crack tip zone \footnote{ The viscosities measured in \cite{HRM}
are of order $10^{8}$ Pa sec. This order of magnitude, huge as
compared with what we expect here for $\eta_{eff}$, must clearly be
assigned to the stress range  which they investigate.
Indeed, far below the yield stress level ($\sigma \ll \sigma^{*}$), thermally
activated CL creep is necessarily extremely slow.}. So, though
encouraging, this comparison is of merely indicative value.\\

Finally, let us come back to the results from series A (same history and
gelatin content, various glycerol contents $\phi$). A power law fit
of the data shown on Figure~\ref{fig:Gamma(mu)A} yields $\Gamma \sim (\mu^{*})^{1.2}$.
Here again, we must conclude that an increase in $\phi$ gives rise
to an increase, not only of the gel stiffness, but also of the
effective viscosity $\eta_{eff}$. Following JHAD, an increased
stiffness means an increase of $c_{hel}$, which signals a change of
solvent quality. In the unstressed gel, this most probably influences
the CL average length as well as the helix fraction. Since changing the
Flory interaction parameter shifts helix-coil transitions, it is
likely to also affect the structural changes shown
by Courty {\it et al} \cite{Courty} to result in large variations of optical activity in
the large strain regime. We expect the value of $\eta_{eff}$ to be
sensitive to these structural modifications.\\

In conclusion, we contend here that fracture of chemical and physical gels
is controlled by different mechanisms~:

  -- stretched chain scission (chemical gels).

  -- viscoplastic cross-link yield leading to chain pull-out (physical
  gels).

Of course, the model formulated here should be tested more completely
by studying crack tip dynamics in other physical hydrogels involving
CL with different structures, such as ionically cross-linked alginates.
More work will also be needed along two directions : (a)
characterization of creep dynamics at larger stress levels than those
used in reference~\cite{HRM}, and of its dependence on solvent viscosity;
(b) more detailed study of slow crack motion, aimed at improving the
reliability of $\mathcal{G}_{0}$-determinations as well as at
testing possible effects of bulk poroelasticity.

\begin{acknowledgement}
We are gratelul to C.Y. Hui for an enlightening discussion. We thank L.
Legrand for his contribution to the analysis of the gel light-scattering
properties.
\end {acknowledgement}


\begin{thebibliography}{100}
%
\bibitem{Clark} A.H. Clark, S.B. Ross-Murphy, Adv. Polymer Sci. {\bf 
83}, 57 (1987).
%
\bibitem{Nij} K. te Nijenhuis,  Adv. Polymer Sci. {\bf 130}, 1 (1997). 
%
\bibitem{HRM}  P. G. Higgs,  S. B. Ross-Murphy, Int. J. Biol. Macromol. {\bf 12},
233 (1990).
%
\bibitem{Vliet} T. van Vliet,  P. Walstra,   Faraday Discuss.  {\bf 101}, 
359 (1995).
%
\bibitem{Lee} K. Y. Lee, D. J. Mooney, Chem. Rev., {\bf 101}, 1869 (2001).
%
\bibitem{Bot} A. Bot, I.A. van Amerongen, R.D. Groot, N.L. Hoekstra, 
W.G.M. Agterof, Polymer Gels and Networks {\bf 4}, 189 (1996). 
%
\bibitem{Evoy} H. Mc Evoy, S. B. Ross-Murphy, A. H.  Clark,  Polymer 
{\bf 26}, 1483 (1985).
%
\bibitem{Bonn}  D. Bonn, H. Kellay, M. Prochnow, K. Ben-Djemiaa, J.  
Meunier, Science {\bf 280},  265  (1998).  
%
\bibitem{Tanaka} Y. Tanaka, K. Fukuao, Y. Miyamoto, Eur. J. Phys. E 
{\bf 3}, 395 (2000). 
%
\bibitem{Mooney} H. J. Kong,  E. Wong,  D. J. Mooney,
Macromolecules {\bf 36}, 4582 (2003).
%
\bibitem{Nature} T. Baumberger, C. Caroli, D. Martina, Nature 
Materials, doi:10.1038/nmat1666 (2006).
%
\bibitem{JHAD} C. Joly-Duhamel, D. Hellio, A. Ajdari, M. Djabourov, 
Langmuir {\bf 18}, 7158 (2002). 
%
\bibitem{Ferry} J.-L. Laurent, P. A. Janmey, J. D. Ferry, J. Rheol.
{\bf 24}, 87 (1980).
\bibitem{DLJ} D.L. Johnson, J. Chem. Phys. {\bf 77}, 1531 (1982). 
%
\bibitem{THB} T. Tanaka, L.O. Hocker, G. B. Benedek, 
J. Chem. Phys. {\bf 59}, 5151 (1973).
%
\bibitem{Ronsin} T. Baumberger, C. Caroli,  O. Ronsin, Eur. Phys. J. 
E{\bf 11}, 85 (2003).
%
\bibitem{Rivlin} R. S. Rivlin, A. G.  Thomas, J. Polymer Sci. {\bf 10}, 
291  (1953). 
%
\bibitem{Rud} J.W. Rudnicki, T-C Hsu, J. Geophys. Res. B {\bf 93}, 
3275 (1988). 
%
\bibitem{Ruina} A. Ruina, M. Sc. Thesis, Brown University, 1978. 
(Can be downloaded from http://ruina.tam.cornell.edu).
%
\bibitem{LT} G. J. Lake, A.G.  Thomas, Proc. R. Soc. London A {\bf 300}, 
108 (1967).
%
\bibitem{Knauss} H.K. Mueller, W.G. Knauss, Trans. Soc. Rheol. {\bf 
15}, 217 (1971).
%
\bibitem{GS} A.N. Gent, J.  Schultz, J. Adhesion {\bf 3}, 281 (1972). 
%
\bibitem{Gent} A. N. Gent, Langmuir {\bf 12}, 4492 (1996). 
%
\bibitem{PGG} P.G. de Gennes, C. R. Acad. Sci. Paris, II {\bf 307}, 
1949 (1988). 
%
\bibitem{HXK} C.-Y. Hui, D. B. Xu, E.J. Kramer, J. Appl. Phys. {\bf 
72} 3294 (1992).
%
\bibitem{BB}, E.A. Brenner, Phys. Rev. E {\bf 71} 
036123 (2005). 
%
\bibitem{Ball} P. G. Higgs, R. C.   Ball,  Macromolecules  {\bf 22}, 
2432 (1989).
%
\bibitem{Nishi} K. Nishinari, S. Koide, K.  Ogino, J. 
Physique {\bf 46}, 793 (1985). 
%
\bibitem{Bo}  B.N.J. Persson, J. Chem. Phys. {\bf 110}, 9713 (1999). 
%
\bibitem{Blunt} C.-Y.Hui, A. Jagota, S. J. Bennison, J. D. Londono,  
Proc. R. Soc. London A {\bf 459}, 1489 (2003).
%
\bibitem{Lawn}  B. R. Lawn, Fracture of brittle solids --- 2nd edn, 
Cambridge, University Press (1993). 
%
\bibitem{Courty} S. Courty, J.L. Gornall, E.M. Terentjev, Biophys. 
J. {\bf 90}, 1019 (2006).
%
\bibitem{Elie} E. Rapha\"el,  P.G. de Gennes, J. Phys. Chem. {\bf 
96}, 4002 (1992). 
%
\bibitem{Fager} L.O. Fager, J.L. Bassani, C.-Y. Hui, D.B. Xu, Int. J. 
Fract. {\bf 52}, 119 (1991). 
\end{thebibliography}
\end{document}